\documentclass[prl,aps,superscriptaddress,floatfix,tightenlines,showpacs,twocolumn]{revtex4-2}
\usepackage{graphicx}
\usepackage{dcolumn}   
\usepackage{bm}        
\usepackage{amssymb}   
\usepackage{amsmath}
\usepackage{url}
\usepackage{layout}
\usepackage{color}
\usepackage{epsfig}
\usepackage{graphicx}
\usepackage{mathrsfs}
\usepackage{bm}
\usepackage{appendix}
\usepackage{color}

\usepackage[thinlines]{easytable}
\usepackage{makecell}
\usepackage{tikz,pgf}
\usepackage{booktabs}
\usepackage{float}
\usepackage{hyperref}
\hypersetup{colorlinks,allcolors=blue}

\def\be{\begin{equation}}       \def\ee{\end{equation}}
\def\bea{\begin{eqnarray}}      \def\eea{\end{eqnarray}}
\def\bp{\begin{pmatrix}} \def\ep{\end{pmatrix}}
\def\beaa{\begin{equation}\begin{aligned}}
		\def\eeaa{\end{aligned}\end{equation}}

\begin{document}

\title{Octupolar Weyl Superconductivity from Electron-electron Interaction}

\author{Zhiming Pan}
\thanks{These two authors contributed equally to this work.}
\affiliation{Department of Physics, Xiamen University, Xiamen 361005, Fujian, China}
\author{Chen Lu}
\thanks{These two authors contributed equally to this work.}
\affiliation{School of Physics, Hangzhou Normal University, Hangzhou 311121, China}
\author{Fan Yang}
\email{yangfan\_blg@bit.edu.cn}
\affiliation{School of Physics, Beijing Institute of Technology, Beijing 100081, China}
\author{Congjun Wu}
\email{wucongjun@westlake.edu.cn}
\affiliation{New Cornerstone Science Laboratory, Department of Physics, School of Science, Westlake University, Hangzhou 310024, Zhejiang, China}
\affiliation{Institute for Theoretical Sciences, Westlake University, Hangzhou 310024, Zhejiang, China}
\affiliation{Key Laboratory for Quantum Materials of Zhejiang Province, School of Science, Westlake University, Hangzhou 310024, Zhejiang, China}
\affiliation{Institute of Natural Sciences, Westlake Institute for Advanced Study, Hangzhou 310024, Zhejiang, China}

\begin{abstract}
Unconventional superconductivity arising from electron-electron interaction can manifest exotic symmetry and topological properties.
We investigate the superconducting pairing symmetry problem based on the 3D cubic $O_h$ symmetry with both weak- and strong-coupling approaches. 
The dominant pairing symmetries belong to the two-dimensional $E_g$ representation at low and intermediate doping levels, and the complex mixing gap function of the
$d_{3z^2-r^2}+id_{x^2-y^2}$-type is energetically favored in the ground state.
Cooper pairs with such a symmetry 
do not possess orbital angular momentum (OAM) moments, which is different from other time-reversal symmetry breaking pairings such as  $p_x+ip_y$ (e.g $^3$He-A) and $d_{x^2-y^2}+id_{xy}$ under the planar hexagonal symmetry.
Instead, they develop the octupolar $O_{xyz}$ component of OAM, which results in 8 nodal points along the body diagonal directions exhibiting an alternating distribution of monopole charges $\pm 1$.
This leads to an intriguing 3D Weyl topological SC, which accommodates nontrivial surface states of Majorana arcs. 
Our results appeal for material realizations and experimental tests in optical lattices.
\end{abstract}
\maketitle

\paragraph{{\color{blue}Introduction}}
The seeking for topological superconductor (TSC) has caught enduring interests, partly due to its relation to topological quantum computation \cite{kitaev2003fault,nayak2008non,sarma2015majorana}. 
Much of the focus has been on fully gapped TSCs \cite{hasan2010colloquium,qi2011topological}, 
including the $p$-wave Kitaev chain \cite{kitaev2001unpaired} and two-dimensional  (2D) chiral superconductors of $(p_x+ip_y)$ and $(d_{x^2-y^2}+id_{xy})$ symmetries \cite{mackenzie2003superconductivity,honerkamp2003instabilities,ogata2003superconducting,nandkishore2012chiral}.
A comprehensive classification of gapped TSCs has been developed, within the ten-fold Altland-Zirnbauer classification \cite{altland1997nonstandard,kitaev2009periodic,ryu2010topological}, which categorizes these systems based on their symmetry and dimensionality.
In three dimensions (3D), there exists a particular type of TSC exhibiting point gap nodes, {\it i.e.}, the Weyl TSC \cite{li2012jtriplet}, which is a superconducting analogy to the Weyl semimetal ~\cite{wan2011pyrochlore,
armitage2018weyl}. 
The concepts of Weyl points are extended to Bogoliubov quasiparticles in superconductors ~\cite{li2012jtriplet, meng2012weyl,sau2012topologically,yang2014dirac,fischer2014chiral,bednik2015superconductivity,volovik2017dirac,yuan2017superconductivity,okugawa2018generic,li2018topological,sumita2018unconventional,roy2019topological,hayes2020weyl,nakai2020weyl,hori2024weyl}. 
The Weyl TSCs not only inherit the non-trivial topological characteristics of Weyl semimetals 
but also exhibit unique superconducting properties, including gapless surface Majorana modes.
Typically the Weyl semimetal or the Weyl TSC except the $^3$He-A phase appears in complicated compounds with spin-orbit coupling (SOC), which is a single-particle band effect from relativistic Dirac equation \cite{manchon2015new,sato2017topological}. 

On the other front, unconventional superconductivity (SC) from many-body interactions is of considerable interests, particularly due to its connection to the high-$T_c$ SC in cuprates ~\cite{bednorz1986cuprate,anderson1987resonating,emery1987theory,dagotto1994correlated,lee2006doping,scalapino2012common}.
The Hubbard model~\cite{hubbard1963electron} is well-known as the simplest lattice model for interacting electrons only with the onsite interaction energy $U$, and plays an important role in the study of strong correlation physics \cite{arovas2022hubbard,qin2022hubbard}. 
In 2D, this model exhibits the antiferromagnetic (AFM) order at half-filling in the square lattice\cite{hirsch1985two,hirsch1989antiferromagnetism,imada1998metal,varney2009quantum}.
Upon doping whether it could yield the $d$-wave SC \cite{kotliar1988superexchange,monthoux1991toward,scalapino1995case} is still under intensive debates in numerical simulations \cite{qin2022hubbard}.


The 3D Hubbard model has been successfully simulated in optical lattices ~\cite{shao2024afm3D}, leading to the observation of the AFM order consistent with numerical simulations ~\cite{hirsch1987simulations,scalettar1989phase,kent2005efficient,rohringer2011critical,hart2015observation,song2024extended}. 
This success offers a new platform for studying interaction effects, 
allowing for precise control over the interaction strength and the temperature to explore various quantum phases  \cite{jaksch2005cold,esslinger2010fermi,bohrdt2021exploration}. 
Particularly, this opens avenues for investigating the emergence of SC in doped systems, where long-range AFM order can be suppressed, leading to the formation of superconducting phases mediated by residual short-range AFM fluctuations. 
It is interesting to investigate what types of exotic SCs are allowed by different 3D point groups, with a focus on the topological aspect.

In this article, we explore the pairing nature from interactions according to the cubic symmetry point group $O_h$. 
The plain Hubbard model in the 3D cubic lattice is taken as a prototype.
Weak-coupling analysis is employed as well as the strong-coupling approach, consistently yielding that the doubly-degenerate $(d_{3z^2-r^2},d_{x^2-y^2})$-wave pairing is dominant at low and intermediate doping levels. 
Ginzburg-Landau (G-L) analysis combined with numerical calculations show that the $d+id$ complex gap function with the octupolar $O_{xyz}$ component of orbital angular momentum (OAM) is energetically favored. 
This gap function exhibits 8 nodes 
exhibiting the alternating pattern of monopole charges $\pm 1$ determined by the octupolar symmetries. 
This is a octupolar Weyl TSC hosting nontrivial surface states of topologically protected Fermi arcs. 
Our result does not depend on a particular model, but is general for systems with the cubic symmetries under repulsive interactions.
It opens a door for studying exotic quantum states in 3D from electron-electron interactions. 

\paragraph{{\color{blue}Model and approach}}
The simplest repulsive Hubbard model is adopted as a prototype for illustrating the pairing symmetry problem in the 3D cubic-lattice,
\begin{eqnarray}
H&=& -t \sum_{\langle i,j\rangle \sigma} c_{i\sigma}^{\dagger} c_{j\sigma}
-\mu \sum_i n_i 
+U \sum_i n_{i\uparrow}  n_{i\downarrow},
\end{eqnarray}
where
$n_{i\sigma}=c_{i\sigma}^{\dagger}c_{i\sigma}$ represents the onsite number of fermion with spin $\sigma$; 
$t$ is the hopping integral; $U$ is the on-site Hubbard repulsion;
$\langle i,j\rangle$ represents the nearest-neighbor (NN) bonds; 
the chemical potential $\mu$ controls the fermion filling and the particle-hole symmetry takes place at $\mu=U/2$.
{ We adopt the hole picture, where doping level $\delta$ is defined as the deviation from half-filling: $\delta=1-n\geq 0$.
Here, $n$ represents the average electron filling per site, with $n=1$ indicating half-filling. 
}

At half-filling, the 3D Hubbard model exhibits long-range Neel AFM order in both limits of $U/t\to 0$ (weak coupling) and $U/t \to +\infty$ (strong coupling)  ~\cite{hirsch1987simulations,scalettar1989phase,kent2005efficient,rohringer2011critical}. 
In the weak-coupling regime, the Fermi surface (FS) exhibits perfect nesting at the wavevector $\bm{Q}=(\pi,\pi,\pi)$, leading to the AFM order. 
In the strong-coupling regime, the system is Mott-insulating.
The superexchange interaction is responsible for the low-energy physics as described by the AFM Heisenberg model,
which shows long-range AFM ordering at low temperatures~\cite{anderson1959new,lee2006doping}.

Upon doping, the AFM order is gradually suppressed and completely disappear above a critical doping level. 
However, the short-range AFM fluctuations are still strong which could prove the pairing glue for SC. 
In the weak-coupling regime, a pair of fermions with opposite momenta acquire an effective attraction via exchanging spin fluctuations as well captured in the
random-phase-approximation (RPA) approach \cite{scalapino1986d,scalapino2012common,takimoto2004strong,yada2005origin,kubo2007pairing,kuroki2008unconventional,graser2009near,liu2013d+,liu2018chiral,liu2023s,zhang2024s}. 
On other hand, in the strong-coupling regime, the exchange interaction leads to the local pairing of the 
resonant-valence-bond type, which upon doping develops phase coherence, leading into SC ~\cite{anderson1987resonating,lee2006doping}. 
Such a picture can be captured in the 
slave-boson-mean-field theory (SBMFT) ~\cite{kotliar1988superexchange,lee2006doping} treatment on the effective $t$-$J$ model. 
Details of the RPA and SBMFT approaches are provided in the Supplementary Materials (S.M.) Sec.~B and C \cite{supp}.

\paragraph{{\color{blue} Weak coupling analysis on pairing symmetry }}
The pairing symmetries on the cubic lattice are classified according to the IRRPs of the $O_h$ group ~\cite{volovik1984unusual,volovik1985superconducting,sigrist1991sc}  see S.M. Sec.~A for details. 
In particular, the $d$-wave channel on the 3D cubic lattice contains the two-fold degenerate $E_g$ and the three-fold degenerate $T_{2g}$ IRRPs.
In contrast, on the 2D square lattice the $d$-wave symmetry splits to 
non-degenerate $B_{1g}$ ($d_{x^2-y^2}$) and $B_{2g}$ ($d_{xy}$) channels.

\begin{figure}[t!]
\centering
\includegraphics[width=0.95\linewidth]{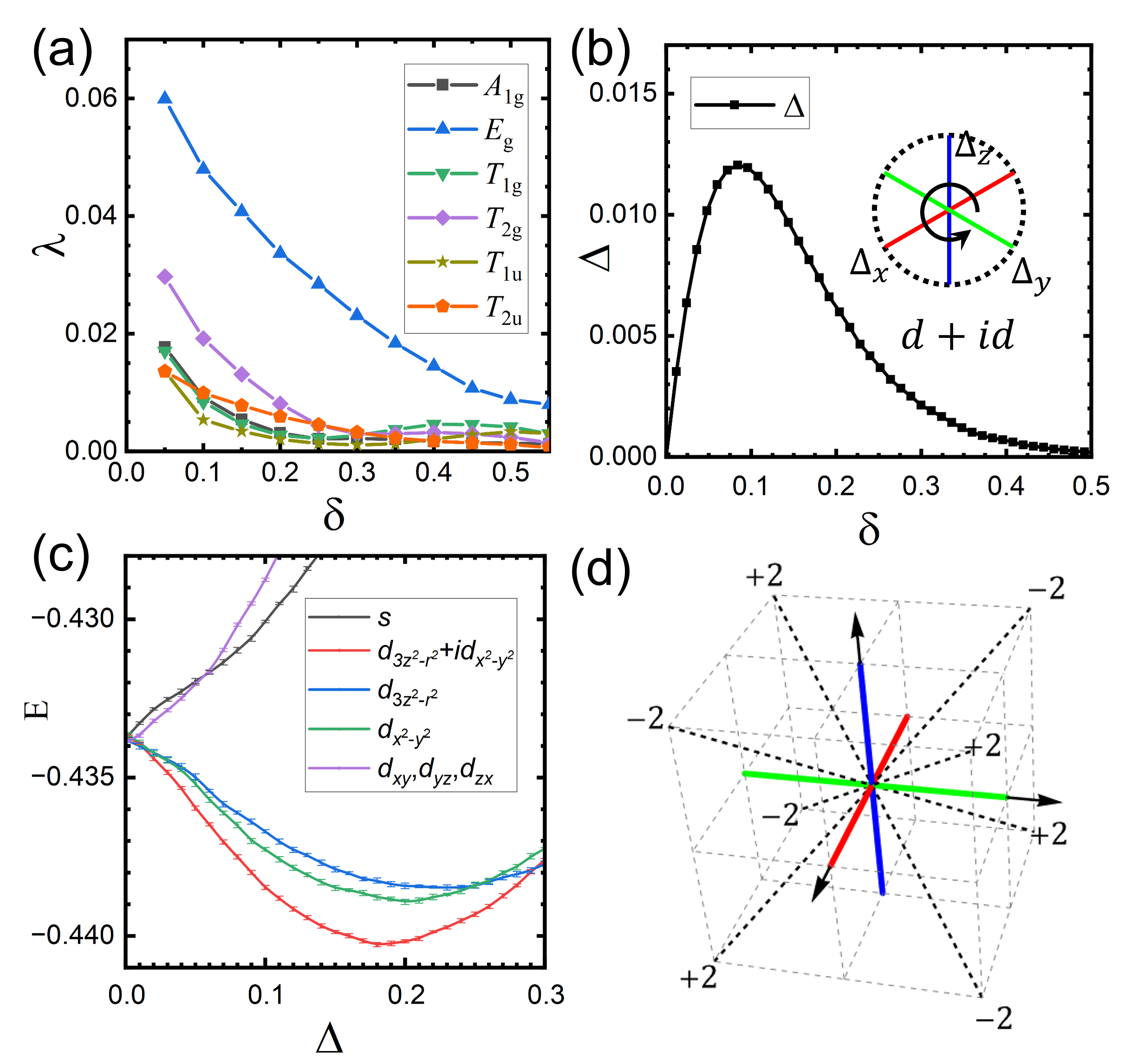}
\caption{{ Simulation of weak-coupling and strong-coupling methods,  consistently showing the dominance of the $E_g$ symmetry pairing.
($a$). RPA analysis of the pairing eigenvalues $\lambda$ versus $\delta$ for various pairing symmetries, calculated with $U/t=1.8$.
($b$). SBMFT calculation of the pairing gap $\Delta$ versus $\delta$ for the $t$-$J$ model, using $J/t=0.4$ ($U/t=10$).
Inset: Illustration of the NN pairing configuration see from $[111]$-direction with the phases of
$1$, $e^{i2\pi/3}$, $e^{i4\pi/3}$ marked as $R$, $G$, $B$, respectively.
($c$). VQMC simulation results showing the energy for different pairing symmetries, using $J/t=0.25$ and $\delta=0.1$.
The results are obtained from minimizing the energy with respect to $\mu$ for various $\Delta$.
($d$). The vorticity projection along body diagonal lines exhibits the alternating pattern of $\pm 2$.
A Cooper pair with such a symmetry carries a non-zero OAM octupolar component $O_{xyz}$.
Consequently, the corresponding gap function exhibits $8$ nodes on the FS. }
}
\label{fig:NumPairings}
\end{figure}

We first analyze the pairing symmetry in the weak coupling regime by comparing the pairing eigenvalues in each symmetry channel. 
The dominant channel is determined by the one with largest eigenvalue.
The RPA approach is adopted to calculate pairing eigenvalues. 
Since the AFM ordering develops near half-filling, we avoid the extremely low doping regime 
in which SC is suppressed, and focus on the low and intermediate hole doping regime $\delta\in(0.05, 0.5)$. 
The numerical results of pairing eigenvalues in various symmetry channels are presented  Fig.~\ref{fig:NumPairings}($a$) as function of $\delta$ for $U/t=1.8$. 
It clearly shows that the $E_g$ IRRP 
dominates in the low doping regime, i.e., the doubly-degenerate
singlet pairing symmetry of $(d_{3z^2-z^2},d_{x^2-y^2})$-wave.  
This pairing symmetry is favored by 
the strong AFM fluctuations in this regime from the nearly nested FS 
~\cite{scalapino1986d,raghu2010sc,ehrlich2020frg}.

The degenerate pairings belonging to a given multi-dimensional IRRP typically will be mixed below $T_c$ due to the non-linearity of the G-L free-energy $\mathcal{F}$. 
Based on the above RPA results, we consider the gap functions $\Delta_{1,2}$ representing the $d_{3z^2-r^2}$ and $d_{x^2-y^2}$ channels, respectively.
Up to the quartic order, $\mathcal{F}$ is constrained by the $U(1)\otimes O_h$ symmetry, hence, takes the following form \cite{sigrist1991sc,agterberg1999conventional},
\begin{equation}
\begin{aligned}
\mathcal{F} &=\alpha \big( |\Delta_1|^2 +|\Delta_2|^2 \big)
+\beta_1 \big( 
|\Delta_1|^2 +|\Delta_2|^2
\big)^2   \\
&+ \beta_2 \big| \Delta_1^* \Delta_2 - \Delta_2^* \Delta_1 \big|^2,
\end{aligned}
\end{equation}
where $\alpha<0$ for setting up the SC state, and $\beta_1>0$ as required by the thermodynamic stability.
The relative phase between $\Delta_1$ and $\Delta_2$ is determined by the sign of $\beta_2$.
For $\beta_2<0$, the free-energy $\mathcal{F}$ is minimized when $\Delta_{1,2}$ exhibit an equal magnitude and the phase difference of $\pm\frac{\pi}{2}$, i.e., 
\begin{align}
\Delta=(\Delta_1,\Delta_2) =\Delta (1,\pm i).
\end{align}
This pairing symmetry is denoted   $d_{3z^2-r^2} \pm i d_{x^2-y^2}$, abbreviated as $d+id$ below, breaking time-reversal symmetry spontaneously. 
On the contrary, if $\beta_2>0$, the free-energy $\mathcal{F}$ favors that $(\Delta_1,\Delta_2)$ share the same phase and their magnitudes 
is determined by higher-order free energy. 
This is an example of nematic pairing symmetry denoted as $d+d$ below. 
See more details of the G-L analysis in S.M. Sec.~A.

The microscopic calculation of the G-L parameters $\beta_2$ would be complicated. 
Nevertheless, typically the weak coupling theory favors the $d+id$ pairing.
The gap function magnitude of this pairing remains the full cubic symmetry of $O_h$ exhibiting nodal points on the FS. 
In contrast, the $d+d$ pairing exhibits an anisotropic gap function with nodal lines. 
Hence, the distribution of $d+id$ gap function over the FS is more uniform than that of the nematic $d+d$ one.
The free energy at the mean-field level is a convex function of the gap function, which favors its uniform distribution over the FS. 
Hence, the $d+id$ pairing should be favored in the weak coupling regime, which spontaneously breaks time-reversal symmetry. 
{We have performed detailed mean-field calculations subsequent to the RPA study, including both complex $d+id$ and real $d+d$ mixed pairing states within the $E_g$ symmetry. 
These calculations show that the complex mixed pairing state is energetically favorable, which implies that, $\beta_2<0$, consistent with the weak coupling theory that favors the $d+id$ pairing. (For details, see S.M. Sect. B).}

Our RPA calculations in the weak-coupling regime also yield the $T_{2g}$ singlet pairing symmetry ($d_{xy},d_{xz},d_{yz})$, or, the triplet $T_{1u}$ symmetry (the $p$-wave) at sufficiently high doping levels $\delta>0.6$. 
The resulting pairing ground states also realize TRSB SC \cite{xu2023frustrated} with exotic properties such as node points and the sextetting order (charge 6e) \cite{agterberg2011conventional,zhou2022chern,pan2024frustrated,zhang2024higgs,ge2024charge}. 
These results are provided in the S.M. Sec.~B.

\paragraph{{\color{blue} Strong coupling analysis on pairing symmetry }}
We next turn to the strong coupling regime, in which the low energy physics is described by the effective $t$-$J$ model,
\begin{equation}
\begin{aligned}
H=-t \sum_{\langle i,j\rangle \sigma} c_{i\sigma}^{\dagger} c_{j\sigma}+J\sum_{\langle i,j\rangle} 
\Big(\bm{S}_i\cdot \bm{S}_j -\frac{1}{4} \Big),
\end{aligned}
\end{equation}
where $\bm{S}_i$ is the spin operator at site $i$ and $J=4t^2/U$. Note that the no-double-occupant constraint is imposed on the Hilbert space. 
The superexchange interaction $J$ induces the NN pairings $\Delta_{x,y,z}$ along the 
$x,y$ and $z$-directions, respectively.
Calculations based on SBMFT 
reveal that the NN pairings 
take the form of $(\Delta_x,\Delta_y,\Delta_z)=\Delta(1, e^{i2\pi/3},e^{ i4\pi/3})$ or its symmetry equivalences. 
This is the same as the $d+id$ symmetry consistent with the RPA results in the weak-coupling regime. 
The doping $\delta$-dependence of the gap function strength $\Delta$ is depicted in Fig.~\ref{fig:NumPairings}($b$), which shows that as increasing doping 
SC quickly emerges and then gradually decreases, exhibiting a dome feature similar to $\lambda$'s dependence on $\delta$ as shown in Fig.~\ref{fig:NumPairings}($a$).   
{ Details of SBMFT approach and further simulated results are provided in S.M. Sect.C.}

{ To corroborate the SBMFT results, variational quantum Monte Carlo (VQMC) simulation were also performed for the $t$-$J$ model (see S.M. Sect.D for details.).
For each candidate pairing symmetry, the ground state energy was optimized with respect to the chemical potential $\mu$,
employing a Gutzwiller-projected BCS trial wavefunction.
Fig.~\ref{fig:NumPairings}(c) summarizes the VQMC energy comparison across various dominant pairing symmetries.
These results consistently demonstrate that the $d+id$ pairing state is energetically the most favorable, thereby providing strong support to its dominance in the strong-coupling regime.
}


\paragraph{{\color{blue} Octupolar 
and Weyl TSC}}
The $d+id$ pairing with the $E_g$ bases is quite different from the $p_x+ip_y$ type of $^3$He-A \cite{leggett1975he3,volovik2003book}, and is also different from
the $d_{x^2-y^2}+i d_{xy}$ type in hexagonal lattices \cite{sigrist1991sc,honerkamp2003instabilities,ogata2003superconducting,nandkishore2012chiral}.
The pairing symmetries of the latter two cases carry the orbital angular momentum (OAM) quantum number with $L_z=1$ and $2$, respectively. 
In contrast, $d_{3z^2-r^2}$ and $d_{x^2-y^2}$ cannot be connected by $L_\pm =L_x\pm iL_y$. 
The expectation values of $L_{x,y,z}$
and its quardrupole tensor components vanish in the $d+id$ state of the $E_g$ version. 
Viewed from the body diagonal directions, the system exhibits the alternating vorticities of $\pm 2$, {\it i.e.},  the octupolar pattern, as illustrated in Fig. \ref{fig:NumPairings} ($d$).
Indeed, a Cooper pair with such a symmetry exhibits a non-zero OAM octupole component defined as $O_{xyz}= \frac{1}{3!} \sum_{P} L_i L_j L_k$ with $P$ representing the permutations of $x,y,z$.

The octupolar pairing symmetry of $d_{3z^2-r^2} \pm i d_{x^2-y^2}$ obtained at the low and intermediate doping regimes exhibits interesting 
symmetry and topological properties.
Without loss of generality, 
its gap function in momentum space is expressed as 
\begin{equation}
\begin{aligned}
\Delta(\bm{k}) =& \Delta \big( \cos k_x 
+e^{i\frac{2\pi}{3}} \cos k_y
+e^{i\frac{4\pi}{3}} \cos k_z\big).
\end{aligned}  
\end{equation}
Its nodal direction appear along the body diagonal ones $(\pm 1, \pm 1, \pm 1)$ in consistent with the octupolar symmetry. 
In comparison, the $p_x+ip_y$ pairing can be viewed as the dipolar symmetry exhibiting nonzero $L_z$ and the node is along the $z$-axis.  
In the Nambu representation, the Bogoliubov-de Gennes Hamilton is given by,
\begin{equation}
\begin{aligned}
H_{\text{BdG}}=& 
\sum_{\bm{k}} \begin{pmatrix}
c_{\bm{k}\uparrow}^{\dagger} & c_{-\bm{k}\downarrow}
\end{pmatrix} H(\bm{k})
\begin{pmatrix}
c_{\bm{k}\uparrow} \\ c_{-\bm{k}\downarrow}^{\dagger} 
\end{pmatrix},  \\
H(\bm{k}) =&\varepsilon(\bm{k}) \tau_3
+\mathrm{Re}\Delta(\bm{k}) \tau_1
-\mathrm{Im}\Delta(\bm{k}) \tau_2,
\end{aligned}
\label{eq:BdGHam}
\end{equation}
where $\varepsilon(\bm{k})=-2t(\cos k_x+\cos k_y +\cos k_z)-\mu$ is the kinetic energy and $\tau_{i}$ is the Pauli matrix in the Nambu space.
The quasi-particle dispersion exhibits 8 nodal points located at the intersections of the body diagonal directions and the FS, i.e., 
$\bm{k}_{\text{node}}=\frac{k_F}{\sqrt{3}}(\pm 1,\pm 1,\pm 1)$, where $k_F$ is the Fermi wavevector. 
Due to TRSB, these node points could be considered as ``Weyl nodes" for the Bogoliubov quasiparticles~\cite{li2012jtriplet,armitage2018weyl}.

\begin{figure}[t!]
\centering
\includegraphics[width=0.9\linewidth]{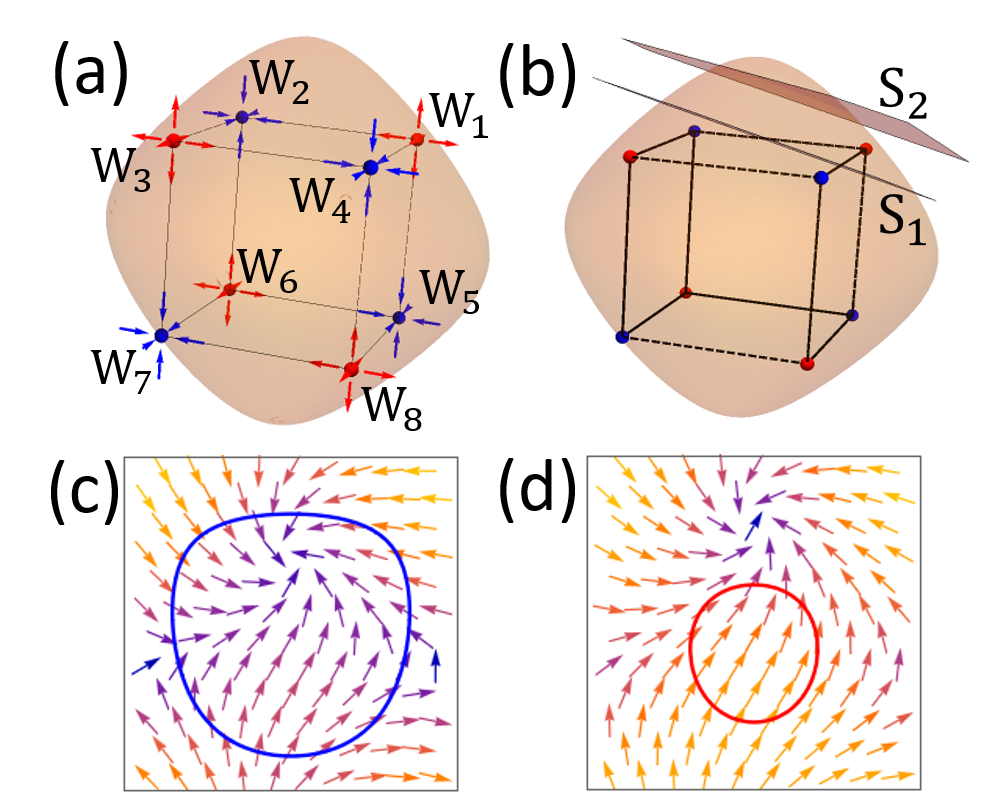}
\caption{
($a$). The 8 Weyl points around the FS with alternating monopole charges exhibiting the octupolar pattern. 
The red (blue) circles indicate Weyl points acting as positive (negative) magnetic monopoles.
($b$). Cross-sections ($S_1$ and $S_2$) with topologically non-trivial and trivial 2D pairing configurations, shown in 
($c$) with $C=1$ for $S_1$ and ($d$) with $C=0$ for $S_2$, respectively. 
Small arrows indicate the orientations of the pairing phase.
}
\label{fig:WeylPoints}
\end{figure}

Near the nodal points, the quasi-particles exhibit the following the linear dispersion relation,
\begin{equation}
\begin{aligned}
H(\bm{p})=& (\bm{v}_1\cdot\bm{p})\tau_1 
+(\bm{v}_2\cdot\bm{p})\tau_2 
+(\bm{v}_3\cdot\bm{p})\tau_3 
\end{aligned}
\end{equation}
where $\bm{p}\equiv \bm{k}-\bm{k}_{\text{node}}$ is the relative momentum with respect to $\bm{k}_{\text{node}}$
and $\bm{v}_i$ ($i=1,2,3$) are unit vectors pointing along the local velocity orientations, see S.M. Sec.~{E} for details.
The 3D chirality (monopole charge) associated with a given Weyl node can be directly calculated as $C=\mathrm{Sgn}\big[\bm{v}_1\cdot \big( \bm{v}_2\times \bm{v}_{3} \big) \big]=\pm 1$
\cite{wan2011pyrochlore,volovik2017dirac}.
The Weyl node at $\bm{k}_{\text{node}}$ with $C=+1 (-1)$ acts as a positive (negative) magnetic monopole in momentum space, and its chirality is determined by the octupolar vorticity along $\bm{k}_{\text{node}}$.
Consequently, the monopole charges of the 8 Weyl nodes are marked in  Fig.~\ref{fig:WeylPoints}($a$), exhibiting 
the octupolar pattern. 
This low-energy effective theory suggests a Weyl superconductor \cite{li2012jtriplet}, which is a superconducting counterpart to the Weyl semimetal.

\begin{figure}[t!]
\centering
\includegraphics[width=0.9\linewidth]{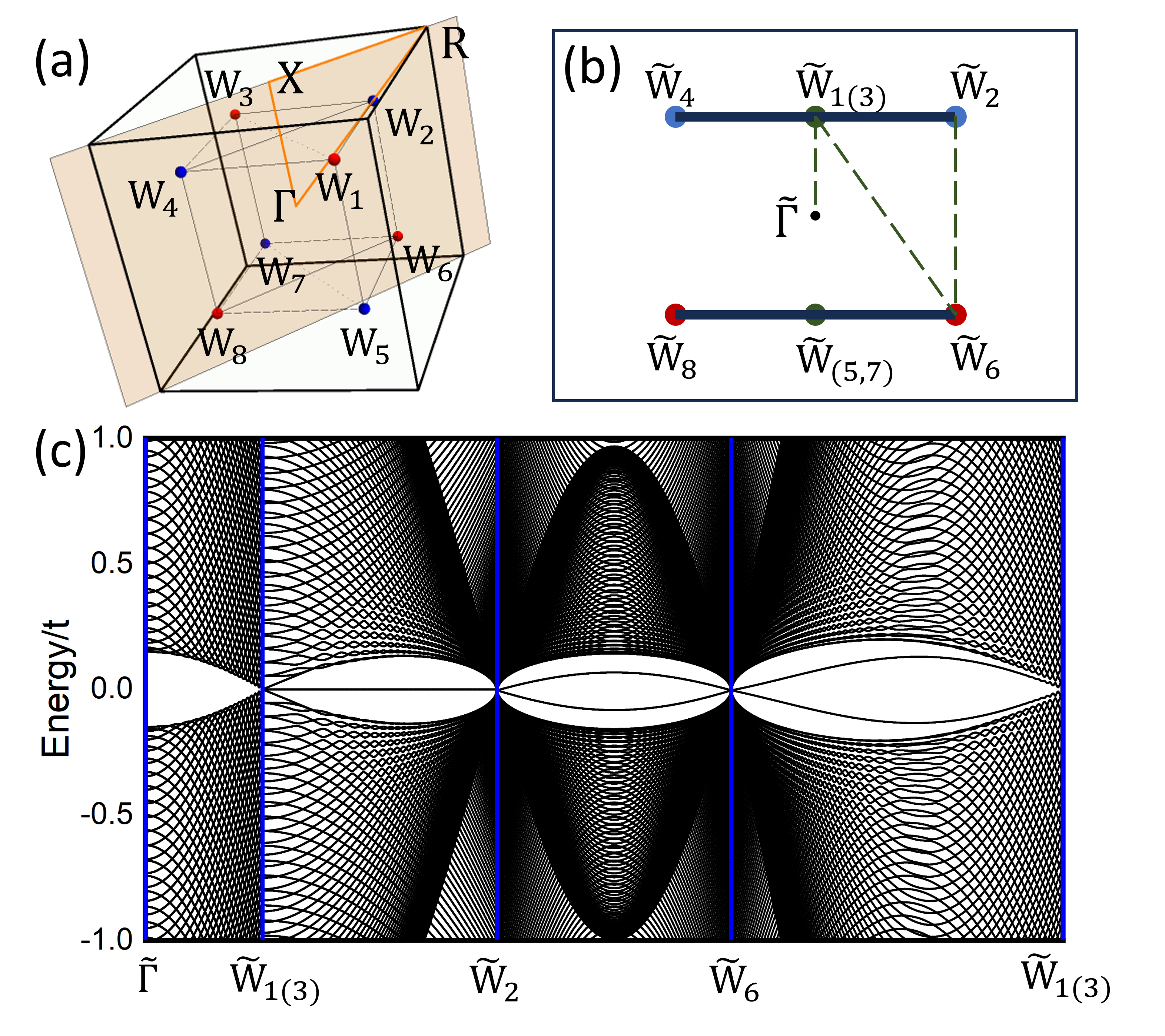}
\caption{($a$). Schematic diagram for the $(110)$-surface (shadow plane). 
($b$). Projected Weyl nodes $\tilde{W}_i$ in the $(110)$ surface BZ and the Fermi arcs (solid lines).  
($c$). Energy spectrum along the path $\tilde{\Gamma}-\tilde{W}_{1(3)}-\tilde{W}_2-\tilde{W}_6-\tilde{W}_{1(3)}$ marked in ($b$) (dashed lines)
for $|\Delta|=0.1t$ and $\mu=-3t$ { based on Eq.~(\ref{eq:BdGHam})}.
}
\label{fig:110SurfaceFermiArc}
\end{figure}

The emergence of Bogoliubov-Weyl points in this pairing state is protected by topological reasons. 
Construct an arbitrary 2D cross-section that intersects the 3D FS while avoiding the Weyl points. 
It results in a 2D gapped superconductor, which can be topologically either trivial or non-trivial depending on the location of the cross-section. 
As illustrated in Fig.~\ref{fig:WeylPoints}($b$),
as the section plane is shifted to cross a Weyl point, the 2D Chern number describing the pairing phase winding around the 2D FS ~\cite{hasan2010colloquium,qi2011topological} changes from $C=1$ before crossing the point ($c$) to $C=0$  afterward ($d$). 
In other words, the 2D SC within the slice evolves from a topological one to a trivial one. 
Hence, an intermediate Weyl node is unavoidable.

With the bulk-boundary correspondence, the Weyl SC exhibits non-trivial Fermi-arc surface states, which are the Majorana arcs of the Bogoliubov quasi-particles.
Specifically, we consider the $(110)$ surface shown in Fig.~\ref{fig:110SurfaceFermiArc}($a$). 
The projection of the 8 Weyl nodes $W_i (i=1 \sim 8)$ onto this surface results in 8 points $\tilde{W}_i$ marked in the surface Brillouin Zone (BZ) shown in Fig.~\ref{fig:110SurfaceFermiArc}($b$). 
To conveniently visualize the special role of the Weyl points in generating gapless surface modes, we calculate the surface spectrum along the path $\tilde{\Gamma}-\tilde{W}_{1(3)}-\tilde{W}_2-\tilde{W}_6-\tilde{W}_{1(3)}$ in the surface BZ (the dashed lines in Fig.~\ref{fig:110SurfaceFermiArc}($b$)) as shown in  Fig.~\ref{fig:110SurfaceFermiArc}($c$). 
Particularly, along the path from $\tilde{W}_{1(3)}$ to $\tilde{W}_2$, there exist a line of gapless modes forming  the Fermi arc $\tilde{W}_{1(3)}-\tilde{W}_2$ exhibited in Fig.~\ref{fig:110SurfaceFermiArc}($b$).  The other Fermi arcs in Fig.~\ref{fig:110SurfaceFermiArc}($b$) are obtained by symmetry. 
Note that the Fermi arcs $\tilde{W}_{1(3)}-\tilde{W}_2$ and $\tilde{W}_{1(3)}-\tilde{W}_4$ are connected at the $\tilde{W}_{1(3)}$ point on the $(110)$ surface.

\paragraph{{\color{blue} Conclusion and Discussion}}
The SC state arising from electron-electron interactions within the context of 3D cubic symmetry is investigated, using the Hubbard model on a 3D cubic lattice as a prototype. 
The results from RPA calculations in the weak coupling regime supplemented by G-L analysis, as well as strong-coupling calculations, reveal the emergence of a TRSB $d+id$ pairing upon doping. 
This state suggests a novel octupolar Weyl TSC, with topological protected Bogoliubov-Weyl nodes and surface Majorana arcs.

The exotic Weyl SC might be realized in certain materials with cubic symmetry or related structure.
Previous studies have suggested that heavy fermion compounds with cubic symmetry, such as UBe$_{13}$ and CeIn$_3$, could exhibit unconventional SC ~\cite{yanase2003theory,pfleiderer2009superconducting}.
{ Similar Weyl SC phases have also been theoretically predicted to emerge in doped 3D spin-$3/2$ Luttinger semimetals ~\cite{roy2019topological}}
Recently, experimental signature of $d$-wave pairing has been proposed in LiTi$_2$O$_4$ \cite{xue2022fourfold}.
Similar $d+id$ state has been suggested for high-$T$ phase cubic CsW$_2$O$_6$ \cite{streltsov2016spin}.
Recent advancements in simulating the AFM phase transition in the 3D Hubbard model \cite{shao2024afm3D} provide a promising pathway for realizing the exotic Weyl SC state proposed in this study upon doping.

Notably, in contrast to 2D systems, which are significantly affected by stronger quantum fluctuations, the AFM order in 3D is more robust \cite{arita1999spin}.
Upon doping, the 3D AFM order will persist over a broader doping range than in its 2D counterpart.
To facilitate the observation of SC, it is crucial to include factors that can suppress the AFM order, such as frustration introduced by next-NN hopping or structural modifications to the lattice.
These factors can be effectively manipulated in optical lattice systems \cite{jaksch2005cold,esslinger2010fermi,bohrdt2021exploration,shao2024afm3D}, offering potential experimental avenues for realizing the predicted Weyl SC state.

{\it Acknowledgments}
We are grateful to Sheng-Shan Qin, Cheng-Cheng Liu and Yan-Xia Xing for stimulating discussions. 
C. W. and F. Y. are supported by the National Natural Science Foundation of China (NSFC) under the Grant No. 12234016, and also supported by the NSFC under the Grant Nos. 12174317 and 12074031, respectively. C.L. is supported by the
National Natural Science Foundation of China under the Grants No. 12304180. This work has been supported by the New Cornerstone Science Foundation.


\twocolumngrid
\bibliography{references}

\newpage

\onecolumngrid
\begin{appendix}

\newpage
\section*{Supplemental Material}

\renewcommand{\theequation}{A\arabic{equation}}
\setcounter{equation}{0}
\renewcommand{\thefigure}{A\arabic{figure}}
\setcounter{figure}{0}
\renewcommand{\thetable}{A\arabic{table}}
\setcounter{table}{0}

\tableofcontents

\renewcommand{\theequation}{A\arabic{equation}}
\setcounter{equation}{0}
\renewcommand{\thefigure}{A\arabic{figure}}
\setcounter{figure}{0}
\renewcommand{\thetable}{A\arabic{table}}
\setcounter{table}{0}

\section{A. Pairing symmetries and Ginzburg-Landau analysis}
The pairing symmetries on the cubic lattice are classified according to the irreducible representations (IRRPs) of the $O_h$ group in Tab.~\ref{tab:GapCubicRepre} ~\cite{volovik1984unusual,volovik1985superconducting,sigrist1991sc}, 
which can be intuitively represented by the lowest order symmetry polynomials. 
The lower index $g$ corresponds to even parity spin singlet pairing symmetries (e.g. $s$ and $d$-waves), and $u$ represents odd parity spin triplet pairing symmetries (e.g. $p$- and $f$- waves).

\begin{table}[b!]
\centering
\begin{tabular}{|c|c|c|c|}
\hline
(even) Rep. & Basis functions & & nodal structure   \\
\hline
$A_{1g}$ & \makecell{$1$ \\ $x^2+y^2+z^2$} &  \makecell{$s$-wave  \\ $s$-wave}  & full-gap \\
\hline
$A_{2g}$ & $(x^2-y^2)(y^2-z^2)(z^2-x^2)$  &  & line-node  \\
\hline
$E_{g}$ & $\{3z^2-r^2,x^2-y^2\}$ & $d$-wave & line- or point-node  \\
\hline
$T_{1g}$ & $\{xy(x^2-y^2),\cdots\}$ & $g$-wave  & line- or point-node  \\
\hline
$T_{2g}$ & $\{xy,yz,zx\}$ & $d$-wave  & full-gap or node \\
\hline
(odd) Rep. & Basic $\bm{d}$-vector functions  &  &  \\
\hline
$A_{1u}$ & $\hat{\bm{x}}k_x+\hat{\bm{y}}k_y+\hat{\bm{z}}k_z$ &  $p$-wave & full-gap  \\
\hline
$A_{2u}$ & $\hat{\bm{x}}k_x(k_z^2-k_y^2)+\cdots$ & $f$-wave & point-node  \\
\hline
$E_{u}$ & $\{2\hat{\bm{z}}k_z-\hat{\bm{x}}k_x-\hat{\bm{y}}k_y,\sqrt{3}(\hat{\bm{x}}k_x-\hat{\bm{y}}k_y)\}$ & $p$-wave  & full-gap or point-node  \\
\hline
$T_{1u}$ & $\{\hat{\bm{y}}k_z-\hat{\bm{z}}k_y,\cdots\}$ & $p$-wave  & point-node  \\
\hline
$T_{2u}$ & $\{\hat{\bm{y}}k_z+\hat{\bm{z}}k_y,\cdots\}$ & $p$-wave & full-gap or point-node  \\
\hline
\end{tabular}
\caption{IRRPs of gap functions for the octahedral group $O_h$, including both the singlet even parity 
and triplet odd parity channels. 
{The nodal structure for each representation is also shown in the last column. 
For multi-component representation, the nodal structure could be full gap, nodal point or nodal line depending on the pairing mixing.}}
\label{tab:GapCubicRepre}
\end{table}

For the two-fold degenerate superconductivity (SC) within the $E_g$ IRRP, $(d_{3z^2-r^2},d_{x^2-y^2})$, 
the Ginzburg-Landau free energy up to the quartic level takes the following form \cite{sigrist1991sc}
\begin{equation}
\begin{aligned}
\mathcal{F} &=\alpha \big( |\Delta_1|^2 +|\Delta_2|^2 \big)
+\beta_1 \big( |\Delta_1|^2 +|\Delta_2|^2 \big)^2 
+ \beta_2 \big| \Delta_1^* \Delta_2 - \Delta_2^* \Delta_1 \big|^2,
\end{aligned}
\end{equation}
where $\Delta_1$, $\Delta_2$ represent the order parameters of $d_{3z^2-r^2}$ and $d_{x^2-y^2}$ pairings, respectively.
For $\beta_2<0$, the phase difference between the two components is $\pm\frac{\pi}{2}$, leading to a complex $1:\pm i$ mixing of the pairing functions with equal amplitude.
This state exhibits spontaneously time-reversal symmetry breaking (TRSB), resulting in Weyl SC.
Conversely, when $\beta_2>0$, the two components share the same phase, resulting in an order parameter of the form:
\begin{align}
\vec{\Delta}=(\Delta_1,\Delta_2) 
\propto (\cos\theta, \sin\theta)
\end{align}
which corresponds to a nematic state that reduces the full rotational symmetry of the $O_h$ group to its subgroup. 
This quartic-order Ginzburg-Landau theory exhibits an accidental continuous degeneracy for the nematic state, which can be lifted through the inclusion of higher-order terms.
To resolve this degeneracy, a sixth-order term respecting the $O_h$ symmetry in the free energy can be added \cite{sigrist1991sc}:
\begin{equation}
\begin{aligned}
\mathcal{F}_{\gamma}
=&\gamma_1 \big( |\Delta_1|^2 +|\Delta_2|^2 \big)^3 
+\gamma_2 \big( |\Delta_1|^2 +|\Delta_2|^2 \big) 
\big| \Delta_1^2 +\Delta_2^2 \big|^2
+\gamma_3 |\Delta_1|^2 \big| 3\Delta_2^2 -\Delta_1^2 \big|^2,
\end{aligned}
\end{equation}
where the $\gamma_3$ term is proportional to $\cos^2(3\theta)$.
It is evident that a positive $\gamma_3$ favors the nematic state with $\theta=(2n+1)\pi/6$ ($n\in\mathbb{Z}$), e.g, $(\Delta_1, \Delta_2) \propto (0, 1)$, 
while a negative $\gamma_3$ supports the state with $\theta=2n\pi/6$ ($n\in\mathbb{Z}$), e.g, $(\Delta_1, \Delta_2) \propto (1,0)$.

Additionally, the $O_h$ symmetry permits the existence of three-fold degenerate SC state, belonging to the $T_{1g}$, $T_{2g}$, $T_{1u}$ or $T_{2u}$ IRRP, as detailed in Tab.~\ref{tab:GapCubicRepre}.
They are characterized by three-component order parameters $\Delta_a$ ($a=1,2,3$).
Up to quartic order, the Ginzburg-Landau free energy for these states is expressed as follows \cite{sigrist1991sc},
\begin{equation}
\begin{aligned}
&\mathcal{F}=\alpha \sum_{a} |\Delta_a|^2 
+\beta_1 \big( \sum_{a} |\Delta_a|^2 \big)^2  
+ \beta_2 \big| \sum_{a} \Delta_a^2 \big|^2
+\beta_3 \Big( |\Delta_1|^2 |\Delta_2|^2 
+|\Delta_2|^2 |\Delta_3|^2 
+|\Delta_3|^2 |\Delta_1|^2 \Big),
\end{aligned}
\end{equation}
where we require $\alpha<0$ and $\beta_1>0$ in the SC state.
The sign of $\beta_2$ is crucial in determining the configuration of the SC state: 
a positive $\beta_2 > 0$ favors a complex mixing structure among the components, 
where a negative $\beta_2<0$ favors a real mixing.

\begin{table}[http!]
\centering
\begin{tabular}{|c|c|c|c|c|}
\hline
$(\Delta_1,\Delta_2,\Delta_3)$  & Free energy & condition & $T_{2g}$ nodal structure & $T_{1u}$ nodal structure   \\
\hline
$e^{i\theta} (1,0,0)$ &  $-4\alpha/(\beta_1+\beta_2)$ & $\beta_3>0,4\beta_2<\beta_3$ & line-node & point-node \\
\hline
$e^{i\theta} (1,1,0)$ &  $-4\alpha/(\beta_1+\beta_2+\beta_3/4)$ & \# & & \\
\hline
$e^{i\theta} (1,i,0)$ &  $-4\alpha/(\beta_1+\beta_3/4)$ & $0<\beta_3<4\beta_2$ & line-node & point-node \\
\hline
$e^{i\theta} (1,1,1)$ &  $-4\alpha/(\beta_1+\beta_2+\beta_3/3)$ & $\beta_2,\beta_3<0$ & full-gap & point-node \\
\hline
$e^{i\theta} (1,\omega,\omega^2)$ &  $-4\alpha/(\beta_1+\beta_3/3)$ & $\beta_3<0<\beta_2$ & point-node & point-node \\
\hline
\end{tabular}
\caption{Possible ground state configurations $(\Delta_1, \Delta_2, \Delta_3)$ and their associated free energy. 
The conditions for each configuration are listed in the third column. 
{The nodal structure corresponding to each state for $T_{2g}$ and $T_{1u}$ IRRRP is provided in the final column.}}
\label{tab:T2gGroundState}
\end{table}

The possible ground states and their associated free energies are summarized in Tab.~\ref{tab:T2gGroundState}.
The ground state is attained when the free energy is minimized \cite{sigrist1991sc}.
Notably, a nematic state featuring only two non-zero components is not favored within the relevant parameter regime. 
For the case where $\beta_3 < 0 < \beta_2$, a TRSB state emerges as the preferred configuration, 
with the order parameter taking the complex structure $(\Delta_1, \Delta_2, \Delta_3) \propto (1, e^{\pm i 2\pi/3}, e^{\pm i 4\pi/3})$.

\section{B. Numerical simulation in the weak-coupling regime and the pairing natures}

{ In the weak-coupling regime, where the perturbative analysis is applicable, 
the random phase approximation (RPA) is a powerful and well-established theoretical framework employed to analyze SC instabilities. 
This approach systematically accounts for pairing mediated by collective electronic excitations, specifically investigating charge and spin fluctuations arising from the particle-hole bubble due to the repulsive Hubbard interaction.

The key quantities in the RPA analysis are the dynamic spin and charge susceptibilities.
The bare (non-interacting, $U=0$) susceptibility $\chi_0(\bm{q}, i\omega_n)$ is calculated using the single-particle band structure $\varepsilon_{\bm{k}}$:
\begin{align*}
\chi_0(\bm{q}, i\omega_n) = \frac{1}{N} \sum_{\bm{k}} \frac{f(\varepsilon_{\bm{k}}) - f(\varepsilon_{\bm{k}+\bm{q}})}{i\omega_n + \varepsilon_{\bm{k}} - \varepsilon_{\bm{k}+\bm{q}} },\qquad
\varepsilon_{\bm{k}} =-2t \Big[ \cos(k_x) +\cos(k_y) +\cos(k_z) \Big] -\mu,
\end{align*}
where, $N$ is the number of lattice site, $t$ is the nearest-neighbor hopping amplitude and $\mu$ is the chemical potential.
$f(\varepsilon)$ denotes the Fermi-Dirac distribution function, 
and $\omega=(2\pi+1)n/\beta$ ($\beta=1/k_BT$, $n\in\mathbb{Z}$) are fermionic Matsubara frequencies.
Within RPA, the charge susceptibility $\chi^{(c)}(\bm{q}, i\omega_n)$, which describes the system's response to charge density fluctuations,
is obtained by summing the geometric series of bubble diagrams connected by the Hubbard interaction $U$,
\begin{align*}
\chi^{(c)}(\bm{q}, i\omega_n) = \frac{\chi_0(\bm{q}, i\omega_n)}{1 + U \chi_0(\bm{q}, i\omega_n)},\quad
\chi^{(c)}(\bm{q})\equiv \chi^{(c)}(\bm{q}, i\omega_n\rightarrow 0).
\end{align*}
Here, we have also introduced static susceptibility $\chi^{(c)}(\bm{q})$ in the $i\omega_n\rightarrow 0$ limit.
The positive sign in the denominator reflects the repulsive nature of the direct Coulomb interaction for charge fluctuations.
Similarly, the spin susceptibility $\chi^{(s)}$ is given by,
\begin{align*}
\chi^{(s)}(\bm{q}, i\omega_n) = \frac{\chi_0(\bm{q}, i\omega_n)}{1 - U \chi_0(\bm{q}, i\omega_n)},\quad
\chi^{(s)}(\bm{q})\equiv \chi^{(s)}(\bm{q}, i\omega_n\rightarrow 0).
\end{align*}
These collective spin and charge fluctuations mediate an effective pairing interaction $V_{\text{eff}}(\bm{k},\bm{q})$ between electrons forming Cooper pairs,
 as described by the pairing term in the Hamiltonian,
\begin{align}
H_{\text{pairing}}= \sum_{\bm{k},\bm{q} \in \text{F.S.S}} V_{\text{eff}}(\bm{k},\bm{q}) 
c_{\bm{k}\uparrow}^{\dagger} c_{-\bm{k}\downarrow}^{\dagger} 
c_{-\bm{q}\downarrow} c_{\bm{q}\uparrow},
\label{eqApp:VeffPairing}
\end{align}
where $\text{F.S.S}$ denotes a small energy shell windows around the Fermi surface and the main contribution of pairing is dominated from the modes near the Fermi surface.
The effective pairing interaction can be decomposed into the singlet $V_{\text{eff}}^{(s)}$ and triplet $V_{\text{eff}}^{(t)}$ channels,
which are related to the dynamic charge/spin susceptibilities as follows:
\begin{equation}
\begin{aligned}
V_{\text{eff}}^{(s)}(\bm{k},\bm{q}) =& U + \frac{3U^2}{4} 
\Big[ \chi^{(s)}(\bm{k}+\bm{q}) +\chi^{(s)}(\bm{k}-\bm{q}) \Big]
- \frac{U^2}{4} \Big[ \chi^{(c)}(\bm{k}+\bm{q}) +\chi^{(c)}(\bm{k}-\bm{q}) \Big], \\
V_{\text{eff}}^{(t)}(\bm{k},\bm{q}) =& - \frac{U^2}{2} 
\Big[ \chi^{(s)}(\bm{k}+\bm{q}) +\chi^{(s)}(\bm{k}-\bm{q}) \Big]
- \frac{U^2}{2} \Big[ \chi^{(c)}(\bm{k}+\bm{q}) +\chi^{(c)}(\bm{k}-\bm{q}) \Big]
\end{aligned}
\label{eqApp:VeffRPA}
\end{equation}
From the effective pairing interaction Eq.~\ref{eqApp:VeffPairing}, a self-consistent gap equation can be derived using a mean-field analysis analogous to conventional BCS theory.
The superconducting gap function is determined by,
\begin{equation}
\tilde{\Delta}(\bm{k}) \equiv 
\sum_{\bm{k},\bm{q}\in \text{F.S.S}} V_{\text{eff}}(\bm{k},\bm{q})\langle c_{-\bm{q}\downarrow} c_{\bm{q}\uparrow}\rangle
= -\sum_{\bm{q}\in \text{F.S.S}} V_{\text{eff}}(\bm{k},\bm{q})
\frac{\tilde{\Delta}(\bm{q})}{E_{\bm{k}}}
\tanh\Big( \frac{\beta}{2}E_{\bm{k}}\Big),
\label{eq:gap_equation}
\end{equation}
where $E_{\bm{k}}=\sqrt{\varepsilon_{\bm{k}}^2+|\tilde{\Delta}(\bm{q})|^2}$ is the Bogoliubov quasiparticle energy.
Near the critical temperature $T_c$, the gap approaches zero,
allowing the gap equation to be linearized,
\begin{align*}
\lambda \tilde{\Delta}(\bm{k})
=-\sum_{\bm{q}\in \text{F.S.S}} V_{\text{eff}}(\bm{k},\bm{q}) \tilde{\Delta}(\bm{q}),\qquad
T_c\propto e^{-1/\lambda}.
\end{align*}
This is an eigenvalue equation where the summation over $\bm{q}$ is restricted to momenta near the Fermi surface within an energy window $\Delta E$.
The dimensionless eigenvalue $\lambda$ determines the pairing strength.
The critical temperature $T_c$ exhibits a BCS-like dependence on the eigenvalue, $T_c\propto e^{-1/\lambda}$.
Therefore, the pairing symmetry channel associated with the maximal eigenvalue $\lambda_{\text{max}}$ corresponds to the largest $T_c$ and represents the dominant pairing instability.
By numerically solving the RPA equations for the susceptibilities and calculating the corresponding eigenvalues $\lambda$ for different symmetry channels, 
one can determine the leading pairing channel. 
}

\begin{figure}[t!]
\centering
\includegraphics[width=0.7\linewidth]{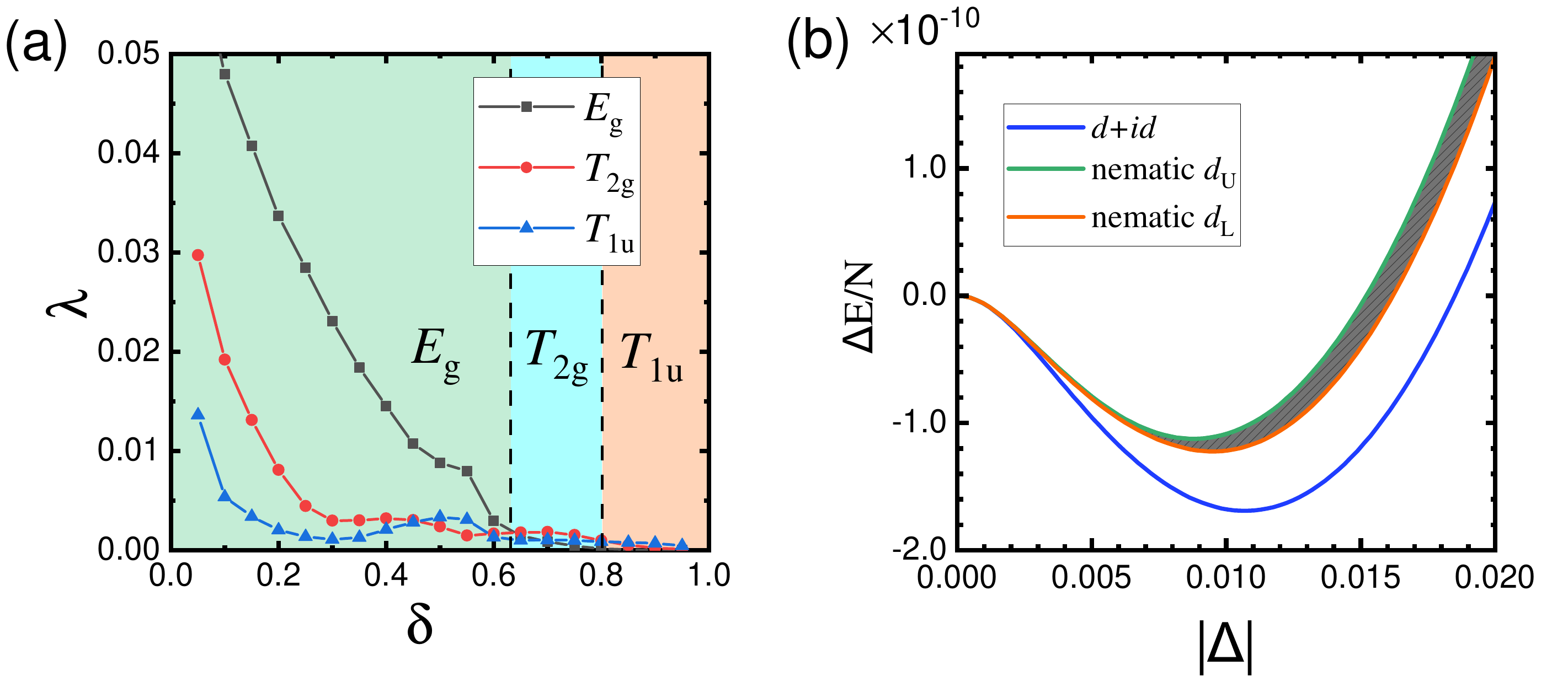}
\caption{(a). Dominated pairing strength eigenvalues $\lambda$ and the phase diagram under doping $\delta$ for $U/t=1.8$.
Near half-filling (lower $\delta$), the SC state is dominated by $d$-wave $E_g$ representation.
As the doping level increases, the SC state evolves into the $d$-wave $T_{2g}$ representation.
At higher doping, the SC state is controlled by a $p$-wave triplet pairing belonging to $T_{1u}$ representation.
{(b). The mean-field energy gains per site, $\Delta E/N$, relative to the normal state for the TRSB complex $d+id$ pairing and the real nematic $d$-wave pairings of the $E_g$ representation from RPA study. 
The parameter values are as follows: doping level $\delta = 0.3$ and $U/t = 3.40$. 
The horizontal axis, $|\Delta|$, represents the rescaled pairing amplitude, and the physical pairing gap around the Fermi surface is given by $\Delta_{\bm{k}}=|\Delta|\cdot f(\bm{k})$, where $f(\bm{k})$ is the normalized angular form factor for each pairing symmetry. 
The pairing gap is simulated within an energy shell $\pm 0.02t$ around the Fermi surface, and the pairing amplitude is normalized within this shell. 
The shaded region corresponds to all possible real nematic pairings with $d_{U,L}$ representing the upper and lower energy bounds of the real nematic pairings. 
The largest amplitude of the form factor $|f(\bm{k})|$ for $d+id$ pairing is given by $1.22 \times 10^{-2}$.}
}
\label{figApp:RPADominatedSCPhase}
\end{figure}

The associated phase diagram for different doping levels $\delta$ based on random-phase approximation (RPA) in the weak-coupling regime is summarized in Fig.~\ref{figApp:RPADominatedSCPhase}(a).
Near half-filling, a two-fold $E_g$ IRRP within $d$-wave pairing $(d_{3z^2-r^2},d_{x^2-y^2})$ is favored.
As the doping $\delta$ increases, the $d$-wave pairing symmetry transitions from $E_g$ IRRP to a three-fold degenerate $T_{2g}$ IRRP $(d_{xy},d_{yz},d_{zx})$ at intermediate doping levels.
At high doping levels, particularly near $\delta=1$ with low electron densities, 
the system exhibits a $p$-wave triplet pairing corresponding to the $T_{1u}$ IRRP.

\begin{figure}[t!]
\centering
\includegraphics[width=0.9\linewidth]{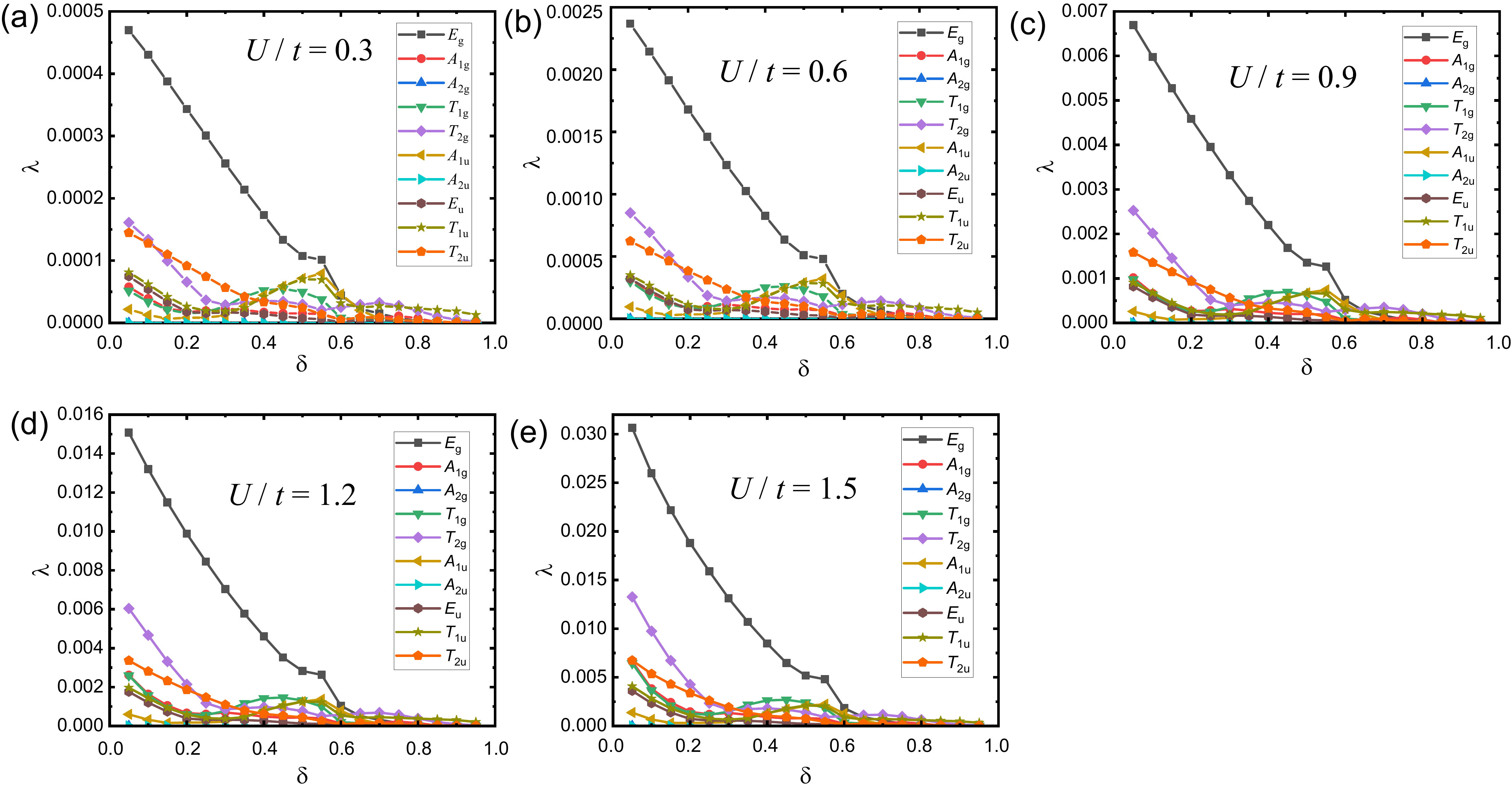}
\caption{(a-e) Pairing eigenvalues for different representation from RPA simulations for varying $U/t=0.3, 0.6, 0.9, 1.2$ and $1.5$ as a function of doping $\delta$. 
The dominant $E_g$ pairing near half-filling (smaller $\delta$) is a consistent feature across these interaction strengths.}
\label{figApp:RPAMultiU}
\end{figure}

{ To further assess the stability of the $E_g$ pairing near half-filling, we performed comprehensive RPA simulations across a range of interaction strengths $U/t$.
As summarized in Fig.~\ref{figApp:RPAMultiU}, the results clearly demonstrate that the dominant pairing instability near half-filling consistently corresponds to the $E_g$ representation for $U/t$ values ranging from $0.3$ to $0.15$.
This consistency across different interaction strengths strongly corroborates our conclusion from the weak-coupling RPA analysis that $E_g$ $d$-wave pairing is the favored state near half-filling.

As depicted in Fig.~\ref{figApp:RPAMultiU}, the results also reveal clear trends for the eigenvalue $\lambda$ associated with the dominant $E_g$ channel:
$\lambda$ generally increases with increasing interaction strength $U/t$ (within the studied range where RPA is applicable). 
This is physically intuitive, as a stronger bare repulsion $U$ can lead to enhanced collective spin (and charge) fluctuations, which in turn provide a more effective pairing ``glue" in appropriate channels.
on the contrary, $\lambda$ decreases with increasing hole doping $\delta$ away from half-filling. 
This behavior likely reflects changes in the Fermi surface geometry and nesting conditions. 
As doping increases, the system may move away from conditions that optimally favor the specific spin fluctuations (e.g., those peaked at or near AFM wavevectors) that most effectively drive the $E_g$ pairing.

With the spin-fluctuation mediated pairing scenario described by RPA, 
the emergence of SC is intimately linked to the leading attractive eigenvalue $\lambda$. 
The SC critical temperature $T_c$ is expected to follow an exponential dependence on this eigenvalue, $T_c\propto e^{-1/\lambda}$.
Consequently, the zero-temperature pairing gap amplitude, $\Delta$, which typically scales with $T_c$ (e.g., $\Delta\approx 1.764k_BT_c$ in BCS theory), will also exhibit similar exponential relation to $\lambda$.
Consequently, the pairing gap $\Delta$ also exhibits a similar exponential dependence on the eigenvalue, $\Delta \propto e^{-1/\lambda}$.
Therefore, the pairing gap $\Delta$ is expected to follow the same trend:
it strengthens with increasing interaction strength $U/t$, and weakens with increasing doping $\delta$.
}

{ The physical intuition for the emergence of an $E_g$ SC state in the 3D cubic-lattice Hubbard model, strongly indicated by the above RPA analysis, draws a compelling analogy to the well-established $d_{x^2-y^2}$-wave pairing in the 2D square-lattice Hubbard model.
Extending this to a 3D cubic system, one might initially consider $d$-wave pairing functions reflecting symmetries along the Cartesian planes, such as $d_{x^2-y^2}$, $d_{y^2-z^2}$, and $d_{z^2-x^2}$.
However, for the classification of pairing states under the cubic ($O_h$) point group, it is essential to use basis functions that transform according to its IRRPs, as depicted in Tab.~\ref{tab:GapCubicRepre}.
The $L=2$ ($d$-wave) spherical harmonics decompose under $O_h$ symmetry into a two-fold degenerate $E_g$ IRRP and a three-fold degenerate $T_{2g}$ IRRP.
The standard choice for the $E_g$ basis functions, which are central to our findings, are $\{d_{3z^2-r^2}, d_{x^2-y^2}\}$.
The $d_{3z^2-r^2}$ function (combined by $d_{y^2-z^2}$ and $d_{z^2-x^2}$) is a specific linear combination that, along with $d_{x^2-y^2}$, correctly spans this two-dimensional $E_g$ space.
The $T_{2g}$ representation, for completeness, is spanned by $\{d_{xy}, d_{yz}, d_{zx}\}$.
Our RPA calculations, which assess the strength of pairing instabilities mediated by electron-electron correlations, consistently identify the $E_g$ channel as the leading instability at low to intermediate doping levels in this 3D Hubbard model.
}

{We perform the Ginzburg-Landau analysis to the pairing configuration based on this two-component $E_g$ orders ($\Delta_1$, $\Delta_2$), with $\Delta_{1,2}$ representing the $d_{3z^2-r^2}$ and $d_{x^2-y^2}$ pairings, respectively. 
Up to the quartic order, two types of configurations are allowed depending on the sign of the coefficient $\beta_2$. 
The contribution from the $\beta_2$ term to the energy is $\beta_2 |\Delta_1 \Delta_2|^2 \sin^2 \Delta\phi$, 
where $\Delta\phi$ is the phase difference between $\Delta_1$ and $\Delta_2$. 
For $\beta_2<0$, to minimize the energy, it gives rise to $\Delta\phi = \pm \frac{\pi}{2}$ and $|\Delta_1|=|\Delta_2|$, which shows that the $d+id$ type complex mixed pairing is favored. 
In contrast, for $\beta_2>0$, it yields $\Delta\phi=0$ leaving the ratio between $|\Delta_1|$, $|\Delta_2|$ arbitrary, which corresponds to a real nematic pairing. 
The degeneracy of the real mixed pairing states will be removed at the 6th order Ginzburg-Landau level analysis (For details, see discussions in S.M. Sec.A).

To further explore these possibilities and identify the energetically favored state, the mean-field calculations subsequent to the RPA study are performed to compare the energies of various pairing states,
including the complex $d+id$ pairing with equal amplitude and $\Delta\phi = \pm \frac{\pi}{2}$, and the real one $\cos\theta d_{3z^2-r^2} + \sin\theta d_{x^2-y^2}$ with $\theta$ controlling the amplitude ratio. 
We calculate the configurations for $\theta$ ranging from $\theta=0$ to $\theta=\pi$, which covers all possible real pairings.

{
Within the mean-field theory, we can study the total energy of the superconducting state by approximating the full pairing Hamiltonian $H$.
The full Hamiltonian consists of the kinetic term and the effective pairing interaction:
\begin{align*}
H=&H_t+H_{\text{pairing}}
=\sum_{\bm{k}\sigma} \varepsilon_{\bm{k}} c_{\bm{k}\sigma}^{\dagger} c_{\bm{k}\sigma} 
+\sum_{\bm{k},\bm{q} \in \text{F.S.S}} V_{\text{eff}}(\bm{k},\bm{q}) 
c_{\bm{k}\uparrow}^{\dagger} c_{-\bm{k}\downarrow}^{\dagger} 
c_{-\bm{q}\downarrow} c_{\bm{k}\uparrow} .
\end{align*}
The interaction term is decoupled via the standard MFT procedure, resulting in the mean-field Hamiltonian $H_{\text{MF}}$:
\begin{align*}
H_{\text{MF}}=&\sum_{\bm{k}\sigma} \varepsilon_{\bm{k}} c_{\bm{k}\sigma}^{\dagger} c_{\bm{k}\sigma}  
+\sum_{\bm{k}} \Big( \Delta_{\bm{k}} c_{\bm{k}\uparrow}^{\dagger} c_{-\bm{k}\downarrow}^{\dagger} +\text{H.c.} \Big)
\end{align*}
The ground-state energy $E$ is then approximated by calculating the expectation value of the original Hamiltonian $H$ using the ground state obtained from $H_{\text{MF}}$:
\begin{align*}
E =\langle H\rangle_{\text{MF}}
=\sum_{\bm{k}\sigma} \varepsilon_{\bm{k}} 
\langle c_{\bm{k}\sigma}^{\dagger} c_{\bm{k}\sigma} \rangle_{\text{MF}}
+\sum_{\bm{k},\bm{q} \in \text{F.S.S}} V_{\text{eff}}(\bm{k},\bm{q}) 
\langle c_{\bm{k}\uparrow}^{\dagger} c_{-\bm{k}\downarrow}^{\dagger}  \rangle_{\text{MF}}
\langle c_{-\bm{q}\downarrow} c_{\bm{q}\uparrow}\rangle_{\text{MF}}
\end{align*} 
under the constraint adjusting the chemical potential implicitly contained within $\varepsilon_{\bm{k}}$,
\begin{align*}
\sum_{\bm{k}\sigma} \langle c_{\bm{k}\sigma}^{\dagger} c_{\bm{k}\sigma} \rangle_{\text{MF}}
=2\big( N_{\varepsilon<\varepsilon_F} -N_{\varepsilon<\varepsilon_{F}-\delta\varepsilon} \big)
\end{align*}
In this equation, $\langle \cdots\rangle_{\text{MF}}$ denotes the expectation value calculated with respect to the mean-field ground state.}
The numerical results, presented in Fig.~\ref{figApp:RPADominatedSCPhase}(b), illustrate the ground-state energy gain per site as a function of the gap amplitudes for the above-mentioned states, based on the results of the RPA calculation. 
It shows that the $d+id$ pairing with equal amplitude and $\Delta\phi = \pm \frac{\pi}{2}$, 
hosts an energy lower than those of all the possible configurations of the real pairings $\cos \theta d_{3z^2-r^2} + \sin \theta d_{x^2-y^2}$, thus is energetically favorable. 
The energy variations among the real mixed states with varying amplitude ratios, which corresponds to the sixth-order terms in the Ginzburg-Landau free energy, are much smaller than the differences between the complex and real mixed states. 
This result is in agreement with both the intuitive arguments presented in the main text and the findings in the strong-coupling limit.
}

At intermediate doping levels, a TRSB superconducting state within the $T_{2g}$ representation may emerge. 
This state corresponds to the following form of the order parameter: 
\begin{align} 
\Delta(\bm{k}) = \frac{\Delta_0}{\sqrt{3}} \big( k_x k_y + e^{i\frac{2\pi}{3}} k_y k_z + e^{i\frac{4\pi}{3}} k_z k_x \big), 
\end{align} 
which exhibits an eight-fold degeneracy linked by the mirror symmetry  or time-reversal transformation. 
In this state, the full $O_h$ point group symmetry is reduced to its subgroup.
The quasi-particle dispersion for this TRSB $T_{2g}$ state features a total of $6+2$ nodal points, denoted as $\bm{k}_{\text{node}}$, 
\begin{align*}
\bm{k}_{\text{node}}
=&k_F(\pm 1,0,0), \quad
k_F(0,\pm 1,0), \quad
k_F(0,0,\pm 1), \quad
\pm k_F( 1, 1, 1).
\end{align*}

For the TRSB superconducting state in the spin-triplet $T_{1u}$ representation, 
the $d$-vector is expressed by: 
\begin{equation}
\begin{aligned}
&d(\bm{k}) \propto\, \hat{\bm{x}} (\omega^2 k_y -\omega k_z)
+\hat{\bm{y}} (k_z -\omega^2 k_x) +\hat{\bm{z}} (\omega k_x -k_y).
\end{aligned}
\end{equation}
with $\omega=e^{i2\pi/3}$.
This TRSB state is non-unitary, exhibiting net spin polarization due to finite $\bm{q}=i\bm{d}\times \bm{d}^*$.
Moreover, it also exhibits nodal points.

Additionally, the three-fold degenerate SC represented by the three components pairing gaps $(\Delta_1,\Delta_2,\Delta_3)$ may realize a sextetting order, 
where six electrons combine to form a bound state, commonly referred to as charge-$6e$ order.
Intriguingly, the $p$-wave triplet representation $T_{1u}$ might realize a sextetting pairing with distinct angular momentum characteristics, differing from the singlet sextetting pairing within the $T_{2g}$ representation.

\section{C. Strong coupling $t$-$J$ model and SBMFT}
In the strong coupling limit, the low-energy effective theory is described by a $t$-$J$ model under the no-double occupancy constraint,
\begin{equation}
\begin{aligned}
H_{tJ} =&-t \sum_{\langle i,j\rangle \sigma} 
\mathcal{P} \big(c_{i\sigma}^{\dagger} c_{j\sigma} +\text{h.c.} \big) \mathcal{P} 
+J \sum_{\langle i,j\rangle} 
\Big(\bm{S}_i\cdot \bm{S}_j -\frac{1}{4} \Big),
\end{aligned}
\end{equation}
where $\bm{S}_i=\frac{1}{2}c_{i}^{\dagger}\bm{\sigma}c_i$ is the spin operator at lattice site $i$ and $J=4t^2/U$ is the nearest-neighbor (NN) antiferromagnetic (AFM) spin superexchange in the large-$U$ limit.
The operator $\mathcal{P}$ projects out double-occupied states, reflecting strong correlation effects.

The constrained $t$-$J$ model can be analyzed using slave-boson mean-field theory (SBMFT)  \cite{kotliar1988superexchange,lee2006doping}.
In this framework, the bare electron operators are expressed in terms of auxiliary holon and spinon operators, maintaining the no-double occupancy constraint:
\begin{align}
c_{i\sigma}^{\dagger} \rightarrow
f_{i\sigma}^{\dagger} h_i,\qquad
c_{i\sigma} \rightarrow
h_i^{\dagger} f_{i\sigma},\qquad
\sum_{\sigma} f_{i\sigma}^{\dagger} f_{i\sigma}
+h_i^{\dagger} h_i=1,
\end{align}
where $h_i^{\dagger}$ and  $f_{i\sigma}^{\dagger}$ are the holon and spinon creation operators at the lattice site $i$, respectively.
In the ground state, the holon condenses, allowing us to replace the holon operator with its condensation density $h_i=h_j^{\dagger}=\sqrt{\delta}$.

Within the mean-field theory, the following spinon hopping and pairing order parameters are introduced,
\begin{equation}
\begin{aligned}
\chi_{\bm{i}\bm{j}}
=&\langle f_{\bm{j}\uparrow}^{\dagger} f_{\bm{i}\uparrow}
+f_{\bm{j}\downarrow}^{\dagger}  f_{\bm{i}\downarrow}\rangle\equiv \chi_{\bm{j}-\bm{i}}, \qquad
\tilde{\Delta}_{\bm{i}\bm{j}}
=\langle f_{\bm{j}\downarrow} f_{\bm{i}\uparrow} 
-f_{\bm{j}\uparrow} f_{\bm{i}\downarrow} \rangle
\equiv \tilde{\Delta}_{\bm{j}-\bm{i}},
\label{eq:OrderPara}
\end{aligned}
\end{equation}
assuming translational symmetry, such that the order parameters depend only on the relative distance between sites. 
We focus on the NN hoppings and pairings along the three directions $\mu=x,y,z$, i.e., $\chi_{\mu}=\chi_{\bm{i},\bm{i}+\hat{\mu}}$ and $\tilde{\Delta}_{\mu}=\tilde{\Delta}_{\bm{i},\bm{i}+\hat{\mu}}$.
The spin-exchange can be decomposed as follows:
\begin{equation}
\begin{aligned}
&J \bm{S}_{i} \cdot \bm{S}_{j}
=-\frac{3}{8} J\Big[\chi_{ij}
\big(f_{i\uparrow}^{\dagger} f_{j\uparrow}
+f_{i\downarrow}^{\dagger} f_{j\downarrow}\big)
+\text{h.c.} \Big]
-\frac{3}{8} J\Big[ \tilde{\Delta}_{ij}
\big(f_{i\uparrow}^{\dagger} f_{j\downarrow}^{\dagger}
-f_{i\downarrow}^{\dagger} f_{j\uparrow}^{\dagger} \big) 
+\text{h.c.} \Big]
+\text{const.}
\end{aligned}
\end{equation}
The slave-boson mean-field Hamiltonian for the spinon field in momentum space is expressed as:
\begin{equation}
\begin{aligned}
&H_{f,\text{MF}}= \sum_{\bm{k} \sigma} \varepsilon(\bm{k})
f_{\bm{k}\sigma}^{\dagger} f_{\bm{k}\sigma}
+\sum_{\bm{k}} \Big[ \tilde{\Delta}(\bm{k}) \big(f_{\bm{k}\uparrow}^{\dagger} f_{-\bm{k}\downarrow}^{\dagger}
-f_{\bm{k}\downarrow}^{\dagger} f_{-\bm{k}\uparrow}^{\dagger} \big) +\text{h.c.} \Big]
\end{aligned}
\end{equation}
where the kinetic energy and spinon pairing field are given by
\begin{align*}
\varepsilon(\bm{k}) =&-2\Big( t\delta+ \frac{3}{8} J \chi \Big) \sum_{\mu} \cos(k_{\mu}) -\mu,\qquad
\tilde{\Delta}(\bm{k}) = -\frac{3}{8}J \sum_{\mu} 
\tilde{\Delta}_{\mu} \cos(k_{\mu}).
\end{align*}
Here, we assume a symmetric hopping fields with $\chi_x=\chi_y=\chi_z\equiv \chi$.
The physical superconducting order parameter is represented as $\Delta\propto \delta \tilde{\Delta}$, where $\delta$ is the holon condensation density.

The pairing configuration $(\Delta_x,\Delta_y,\Delta_z)\propto (1,1,1)$ corresponds to a $s$-wave pairing. 
In contrast, the pairing configuration $(\Delta_x,\Delta_y,\Delta_z)\propto (1,-1,0)$ and $(\Delta_x,\Delta_y,\Delta_z)\propto (-1,-1,2)$ correspond to the $d_{x^2-y^2}$-wave pairing and $d_{3z^2-r^2}$, respectively, both expected to be degenerate under the $E_g$ representation of the $O_h$ symmetry group.

\begin{figure}[t!]
\centering
\includegraphics[width=0.4\linewidth]{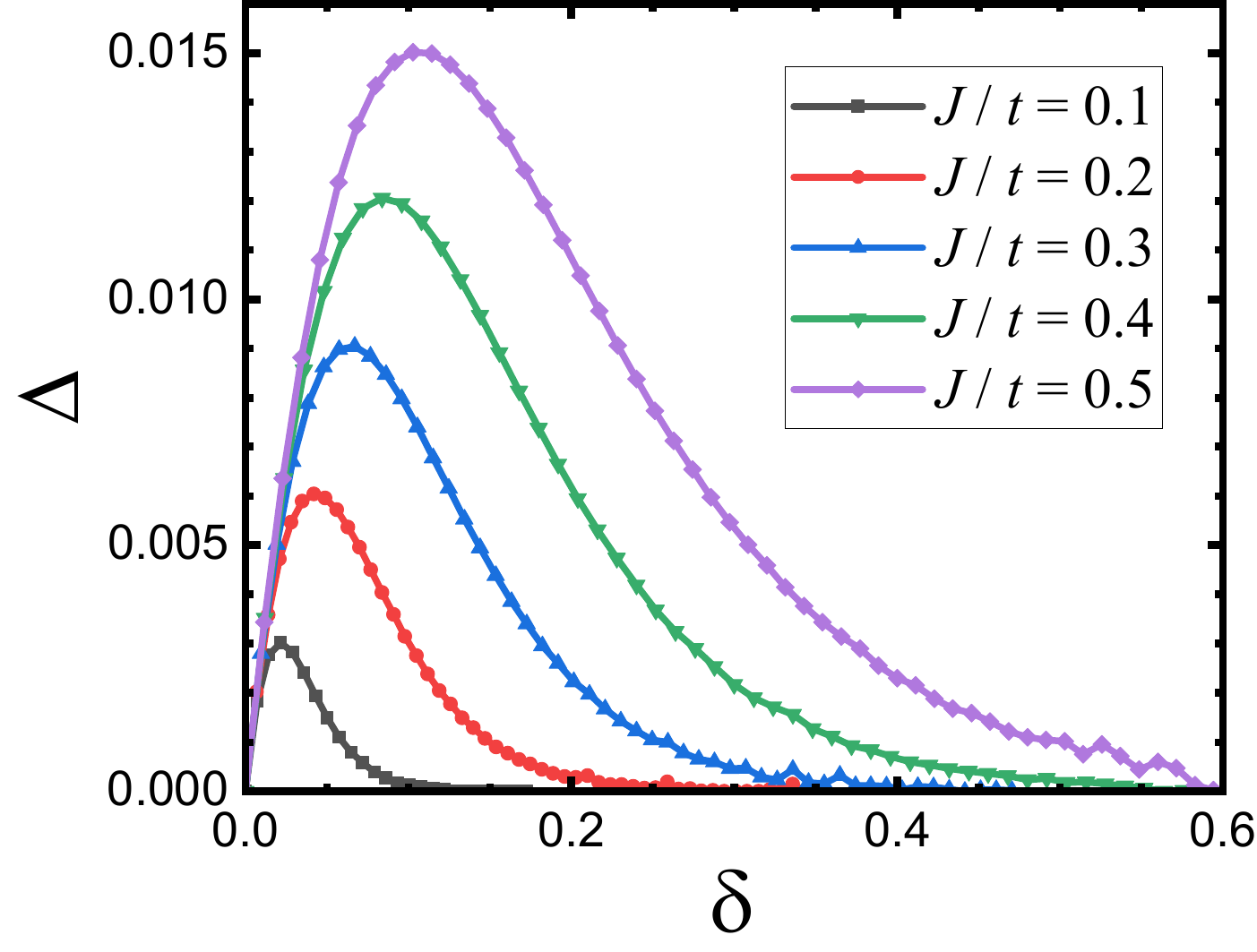}
\caption{Pairing amplitude for the $t$-$J$ model based on the slave-boson mean-field calculation, for a range of $J/t=0.1, 0.2, 0.3, 0.4, 0.5$}
\label{figApp:SBMFmultiJ}
\end{figure}

{
The self-consistent SBMFT gap equations for the nearest-neighbor spinon pairing amplitudes $\tilde{\Delta}_{\mu}$ ($\mu=x,y,z$) are then solved:
\begin{align*}
\tilde{\Delta}_{\mu} =\langle f_{\bm{i}+\hat{\mu}\downarrow} f_{\bm{i}\uparrow} 
-f_{\bm{i}+\hat{\mu}\uparrow} f_{\bm{i}\downarrow} \rangle_{\text{MF}}
=\frac{2}{N} \sum_{\bm{k}} \cos(k_{\mu}) \langle f_{\bm{k}\downarrow} f_{-\bm{k}\uparrow}  \rangle_{\text{MF}}
\end{align*}
where the expectation value $\langle \cdots\rangle_{\text{MF}}$ is calculated with respect to the spinon mean-field Hamiltonian $H_{f,\text{MF}}$.
Within a relevant range of parameter regime (doping levels $\delta$ and exchange coupling $J/t$), 
the numerical results indicate that a $d$-wave superconducting state, characterized by a complex phase structure $(\Delta_x,\Delta_y,\Delta_z)=\delta\tilde{\Delta}_0 (1,e^{\pm i2\pi/3},e^{\pm i4\pi/3})$, is energetically favored.
The resulting overall physical gap amplitude $\Delta_0=\delta\tilde{\Delta}_0$ (where $\delta$ is the hole doping concentration representing the holon condensate density) obtained from these simulations is summarized as a function of doping for various $J/t$ values in Fig.~\ref{figApp:SBMFmultiJ}.
}

{
\section{D. Variational Quantum Monte Carlo simulation}
To further validate the findings from SBMFT in the strong-coupling regime, 
where the Hubbard model maps to the $t$-$J$ model,
we employ Variational Quantum Monte Carlo (VQMC) simulations.

The trial wavefunction for a pairing state is constructed using a Gutzwiller-projected BCS function:
\begin{align*}
|G\rangle 
=\hat{P}_{G} \Big( \sum_{\bm{k}} \frac{v_{\bm{k}}}{u_{\bm{k}}} c_{\bm{k}\uparrow}^{\dagger} 
c_{-\bm{k}\downarrow}^{\dagger} \Big)^{N_e/2} |\text{vac}\rangle,\qquad
\frac{v_{\bm{k}}}{u_{\bm{k}}} =
\frac{\Delta(\bm{k})}{\varepsilon(\bm{k})+\sqrt{\varepsilon(\bm{k})^2 +\Delta(\bm{k})^2}}
\end{align*}
where $\hat{P}_G$ is the Gutzwiller projection operator that enforces the no-double-occupancy constraint at each site and fixes the total particle number $N_e$ for a given hole doping level $\delta$. 
Consequently, the resulting physical Hilbert space allows only three local configurations: 
empty (hole), singly occupied with spin up, or singly occupied with spin down. 
$\varepsilon(\bm{k})$ represents the kinetic energy dispersion of the non-interacting electrons:
\begin{align*}
\varepsilon_{\bm{k}} =-2t \Big[ \cos(k_x) +\cos(k_y) +\cos(k_z) \Big] -\mu,
\end{align*}
The pairing gap function $\Delta(\bm{k})=\Delta f(\bm{k})$ in the trial wavefunction is chosen to correspond to specific IRRPs of the $O_h$ group,
as shown in Tab.~\ref{tab:GapCubicRepre2}.
Here, $f(\bm{k})$ is the form factor incorporating nearest-neighbor pairing terms. 
In the VQMC simulations, the chemical potential $\mu$ is treated as variational parameter, adjusted to optimize the total energy $E=\langle {G}|H_{tJ}|{G}\rangle/\langle {G}|{G}\rangle$ for a fixed pairing amplitude.

\begin{table}[t!]
\centering
\begin{tabular}{|c|c|}
\hline
Pairings & Form factor $f(\bm{k})$   \\
\hline
$s$ & $2\cos(k_z)+2\cos(k_x)+2\cos(k_y)$  \\
\hline
$d_{3z^2-r^2}+id_{x^2-y^2}$ & $2\cos(k_z)+2e^{i2\pi/3}\cos(k_x)+2e^{i4\pi/3}\cos(k_y)$  \\
\hline
$d_{3z^2-r^2}$ &  $2\cos(k_z)-\cos(k_x)-\cos(k_y)$  \\
\hline
$d_{x^2-y^2}$ &  $2\cos(k_x)-2\cos(k_y)$  \\
\hline
$d_{xy},d_{yz},d_{zx}$ & $2\cos(k_x+k_y)+2\cos(k_x-k_y)$  \\
\hline
$p_x,p_y,p_z$ & $2\sin(k_x)$  \\
\hline
\end{tabular}
\caption{Gap functions for the VQMC calculations}
\label{tab:GapCubicRepre2}
\end{table}

The comparative energy results for different pairing symmetries obtained from VQMC are presented in the main text (Fig. 1(c)).
Noticing that antiferromagnetic superexchange $J$-term acts only between nearest neighbors, naturally favoring singlet and more local pairing mechanisms, we incorporate nearest-neighbor (NN) and next-nearest-neighbor (NNN) range terms. 
Other possibilities listed in Tab.~\ref{tab:GapCubicRepre} were also tested but are not depicted, as they resulted in higher energies than the normal state or the $d$ states.

}

\section{E. TRSB $d+id$ Weyl superconductor}

The TRSB Weyl SC pairing characterized by $d_{3z^2-r^2} \pm i d_{x^2-y^2}$ is expressed as follows:
\begin{align*}
&\Delta_{\eta}(\bm{k}) =\Delta_0 \Big( \cos(k_x) 
+e^{i\eta\frac{2\pi}{3}} \cos(k_y)
+e^{i\eta\frac{4\pi}{3}} \cos(k_z)\Big) \\
=&\Delta_0 \Big( \cos(k_x) 
-\frac{1}{2}\cos(k_y) -\frac{1}{2}\cos(k_z)\Big)
+ \Delta_0 i\eta \frac{\sqrt{3}}{2}\Big( \cos(k_y) -\cos(k_z)\Big),
\end{align*}
where $\eta=\pm 1$ corresponds to the two choices of chirality.
This pairing belongs to the $E_g$ irreducible representation of the $O_h$ point group.
The Bogoliubov-de Gennes (BdG) Hamiltonian can be formulated in the two-component Nambu spinor representation as follows:
\begin{align}
H_{\text{BdG}}=& 
\sum_{\bm{k}} \begin{pmatrix}
c_{\bm{k}\uparrow}^{\dagger} & c_{-\bm{k}\downarrow}
\end{pmatrix}
H(\bm{k})
\begin{pmatrix}
c_{\bm{k}\uparrow} \\ c_{-\bm{k}\downarrow}^{\dagger} 
\end{pmatrix},\qquad
H(\bm{k})= \begin{pmatrix}
\varepsilon(\bm{k}) & \Delta(\bm{k}) \\
\Delta^*(\bm{k}) & -\varepsilon(\bm{k})
\end{pmatrix}.
\end{align}
Here, the kinetic energy is given by
\begin{align*}
\varepsilon(\bm{k})
= -t \Big( \cos(k_x) +\cos(k_y) +\cos(k_z) \Big) -\mu
\end{align*}
where $t$ represents the effective hopping integral.
The Hamiltonian matrix can be expressed in a compact form,
\begin{align}
H(\bm{k}) =\varepsilon(\bm{k}) \tau_3
+\mathrm{Re}\Delta(\bm{k}) \tau_1
-\mathrm{Im}\Delta(\bm{k}) \tau_2
\end{align}
with $\tau_i$ ($i=1,2,3$) denoting the Pauli matrices in the Nambu particle-hole space.

The Fermi surface is defined by the condition $\varepsilon(\bm{k}_F)=0$, with the eight node points corresponding to:
\begin{align*}
\bm{k}_{\text{node}} = \frac{1}{\sqrt{3}} k_F ( \pm 1, \pm 1, \pm 1),
\end{align*}
To investigate these nodes further, we focus on the low-energy effective theory around the two points $\bm{k}_{111}=(1,1,1)k_F/\sqrt{3}$ and $\bm{k}_{\bar{1}\bar{1}\bar{1}}=-(1,1,1)k_F/\sqrt{3}$,
\begin{align*}
H_{\eta}(\bm{k}_{111} +\bm{p})
\approx&\Big( \varepsilon(\bm{k}_{111} +\bm{p}) -\varepsilon(\bm{k}_{111}) \Big) \tau_3
+\Delta_0 \sin\frac{k_F}{\sqrt{3}} \Big( -p_x +\frac{1}{2}p_y+\frac{1}{2}p_z \Big) \tau_1
+\eta \Delta_0 \frac{\sqrt{3}}{2} \big( -p_y+p_z \big) \tau_2,   \\
H_{\eta}(\bm{k}_{\bar{1}\bar{1}\bar{1}} +\bm{p})
\approx&\Big( \varepsilon(\bm{k}_{\bar{1}\bar{1}\bar{1}} +\bm{p}) -\varepsilon(\bm{k}_{\bar{1}\bar{1}\bar{1}}) \Big) \tau_3
-\Delta_0 \sin\frac{k_F}{\sqrt{3}} \Big( -p_x +\frac{1}{2}p_y+\frac{1}{2}p_z \Big) \tau_1
-\eta \Delta_0 \frac{\sqrt{3}}{2} \big( -p_y+p_z \big) \tau_2
\end{align*}
where the kinetic part near the Fermi surface expands to:
\begin{align*}
\varepsilon(\bm{k}_{111} +\bm{p}) -\varepsilon(\bm{k}_{111})
\approx \tilde{t} \sin\frac{k_F}{\sqrt{3}} \big( p_x+p_y+p_z \big),\qquad
\varepsilon(\bm{k}_{\bar{1}\bar{1}\bar{1}} +\bm{p}) -\varepsilon(\bm{k}_{\bar{1}\bar{1}\bar{1}})
\approx -\tilde{t} \sin\frac{k_F}{\sqrt{3}} \big( p_x+p_y+p_z \big)
\end{align*}
Using an appropriate rescaling of the coordinates, we can write the effective low-energy Hamiltonian as a Weyl form:
\begin{equation}
\begin{aligned}
H_{\eta}(\bm{k}_{111} +\bm{p})
\approx& + \Big[ (\bm{v}_1\cdot\bm{p})\tau_1  
+\eta (\bm{v}_2\cdot\bm{p})\tau_2 
+(\bm{v}_3\cdot\bm{p})\tau_3 \Big], \\
H_{\eta}(\bm{k}_{\bar{1}\bar{1}\bar{1}} +\bm{p})
\approx& -\Big[ (\bm{v}_1\cdot\bm{p})\tau_1  
+\eta (\bm{v}_2\cdot\bm{p})\tau_2 
+(\bm{v}_3\cdot\bm{p})\tau_3 \Big]
\end{aligned}
\end{equation}
where the three unit directional vectors are defined as:
\begin{align*}
\bm{v}_1= v\sqrt{\frac{2}{3}} \Big(-1, \frac{1}{2}, \frac{1}{2} \Big),\quad
\bm{v}_2= v\eta \frac{1}{\sqrt{2}} \Big(0, -1, 1 \Big)\quad
\bm{v}_3= v\frac{1}{\sqrt{3}} \Big(1, 1, 1 \Big)
\end{align*}
and $v>0$ is the effective fermi velocity.
The magnetic monopole charge in the momentum space associated with these two Weyl points can be calculated as follows:
\begin{align*}
C_{\eta}(\bm{k}_{111}) 
= \mathrm{Sgn}\Big[ \bm{v}_1\cdot \big( \bm{v}_2\times \bm{v}_3 \big) \Big] 
={\eta},\quad
C_{\eta}(\bm{k}_{\bar{1}\bar{1}\bar{1}}) 
= \mathrm{Sgn}\Big[ \bm{v}_1\cdot \big( \bm{v}_2\times \bm{v}_3 \big) \Big] 
={-\eta}.
\end{align*}

\end{appendix}

\end{document}